\documentclass[prl,showpacs,twocolumn]{revtex4}
\usepackage{amsmath}    
\usepackage{verbatim}   
\usepackage{color}      
\usepackage{subfigure}  
\usepackage{hyperref}   
\usepackage{graphics}
\usepackage{epsfig}
\usepackage[english]{babel}
\usepackage{natbib}
\usepackage{fancybox}
\begin{document}

\title{Clonal selection prevents tragedy of the commons when neighbors compete in a rock-paper-scissors game}
  
\author{Jeppe Juul, Kim Sneppen, and Joachim Mathiesen}

\affiliation{University of Copenhagen, Niels Bohr Institute, Blegdamsvej 17, DK-2100 Copenhagen, Denmark}

\date{\today}

\begin{abstract}
The rock-paper-scissors game is a model example of the on-going cyclic turnover typical of many ecosystems, ranging from the terrestrial and aquatic to the microbial. Here we explore the evolution of a rock-paper-scissors system where three species compete for space. The species are allowed to mutate and change the speed by which they invade one another. In the case when all species have similar mutation rates, we observe a perpetual arms race where no single species prevails. When only two species mutate, their aggressions increase indefinitely until the ecosystem collapses and only the non-mutating species survives. Finally we show that when only one species mutates, group selection removes individual predators with the fastest growth rates, causing the growth rate of the species to stabilize. We explain this group selection quantitatively.
\end{abstract}


\maketitle

\textbf{Introduction}
---When multiple individuals depend on a shared and limited resource, it is in the long-term interest of everyone that the resource is rationed to avoid depletion. However, in the short term it is in the interest of each individual to consume resources fast so as to gain a competitive advantage over their more prudent neighbors. Acting rationally to promote their own self-interest, each individual therefore increases consumption until the resource is depleted \cite{Hardin, Dionisio}. This dilemma, known as the tragedy of the commons, illustrates the need for restrictions on the use of limited resources to ensure sustainable development. The potential consequences of not regulating common property are particularly evident when it comes to issues such as overfishing and global warming. Surprisingly, even among primitive life-forms, such as bacteria and plants, prudent consumption of shared resources has been observed in large groups of competing individuals \cite{Kerr06, MacLean, Rankin, Heilmann}. These individuals have no means of enforcing common restrictions, so the emergence of restraint must have an evolutionary origin.

\begin{figure}[tb]
    	\centering
            	\includegraphics[width=\columnwidth]{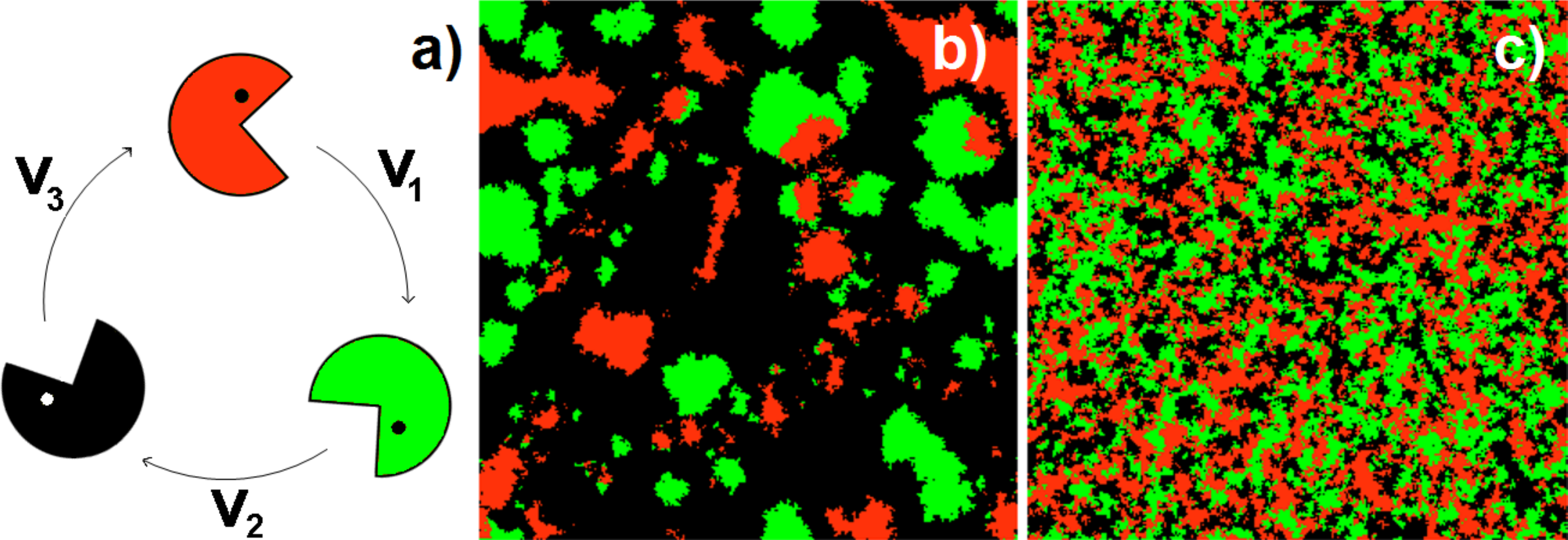}
    	\caption{(Color online)
            	\textbf{a)} We consider a system of three species with cyclic interactions. Species 1 overgrows species 2, which overgrows species 3, which overgrows species 1. Individuals grow at different rates. The mean growth rate for species 1, 2, and 3 are denoted $v_1$, $v_2$, and $v_3$.
            	\textbf{b)} When species 1 grows faster than 2 and 3, species 3 will be more abundant than species 1 and 2. Here $v_2=v_3=v_1/5$.
            	\textbf{c)} When all species grow at the same mean rate, they will be equally abundant. Members of same species will self-organize to form clusters on the lattice.}
    	\label{fig:cycle}
\end{figure}

We here study a community of three species with a cyclic interaction where species 1 overgrows species 2, which overgrows species 3, which, in turn, overgrows species 1 (see fig. \ref{fig:cycle}a). Such intransitive systems, similar to the game rock-paper-scissors, have been identified in numerous ecosystems, ranging from terrestrial and aquatic to microbial ecosystems \cite{Jackson, Taylor, Durrett, Birkhead, Nahum, Lankau, Cameron}.

An interesting property of cyclic interaction is that growing fast will not help a species to gain biomass - it will help its predator! By growing too fast, the species will weaken the population of its prey thereby improving conditions for its predator (see fig. \ref{fig:cycle}b-\ref{fig:cycle}c) \cite{Tainaka, Szabo}. Thus, for the species as a whole it is advantageous to grow slowly, whereas each individual of the species will get a competitive advantage from growing fast. It has been observed, both in simulations and in experiments, that such a species will suffer the tragedy of the commons if they interact globally, but that biodiversity may be maintained if the interactions are local \cite{Reichenbach, Prado, Johnson, Kerr02, Mathiesen, Frean, Boerlijst}. The evolution of restraint must, therefore, depend on the spatial structure of the system. However, a quantitative mechanism for this has yet to be identified.

\textbf{Model}
---In our model, each site of an $L \times L$ square lattice is occupied by a member from one of the species, 1, 2, or 3. Initially, all individuals grow at the same rate. At each time step the following actions take place:
\begin{itemize}
    	\item A random node $i$ and one of its four neighbors $j$ are selected. If $i$ can overgrow $j$ it does so with a probability $v_i$. Hereby, $j$ becomes a member of $i$'s species with the same growth rate $v_j=v_i$
    	\item When $i$ overgrows $j$, it might mutate by a small probability $p_{mutate}$. Hereby, $j$ will either become faster growing than $i$, $v_j = (1+\gamma) v_i$, or slower growing, $v_j = v_i / (1+\gamma)$, where $\gamma$ is a constant.
\end{itemize}
Unless otherwise stated, we have used $L=200$, $\gamma = 0.02$, and $p_{mutate}=5\cdot 10^{-5}$. This choice of parameters ensured, that the range of growth rates within a species would only vary $2-4\%$ at any given time.

\textbf{Results}
---Assume that the species grow at rates $(v_1, v_2, v_3)$ and that the probabilities of finding each at any one lattice site are $(p_1, p_2, p_3)$. In the mean field approximation we achieve a steady state when $v_1 p_1 p_2 = v_2 p_2 p_3 = v_3 p_3 p_1$, leading to \cite{Tainaka, Reichenbach06}
\begin{eqnarray}
    	(p_1,p_2,p_3) = \frac{1}{v_1+v_2+v_3} (v_2,v_3,v_1) \Rightarrow \label{eq:speciesCounts1} \\
    	\frac{p_1}{v_2} = \frac{p_2}{v_3} =\frac{p_3}{v_1}
\label{eq:speciesCounts}
\end{eqnarray}

This steady state is stable for the spatially structured system. In the mean field model, where all sites are neighbors, the abundances of the three species will oscillate with larger and larger amplitudes until biodiversity is lost. It is, therefore, only possible to study the effect of mutations, when the system is spatially structured.

\begin{figure}[tbp]
    	\begin{center}
                    	\includegraphics[width=\columnwidth]{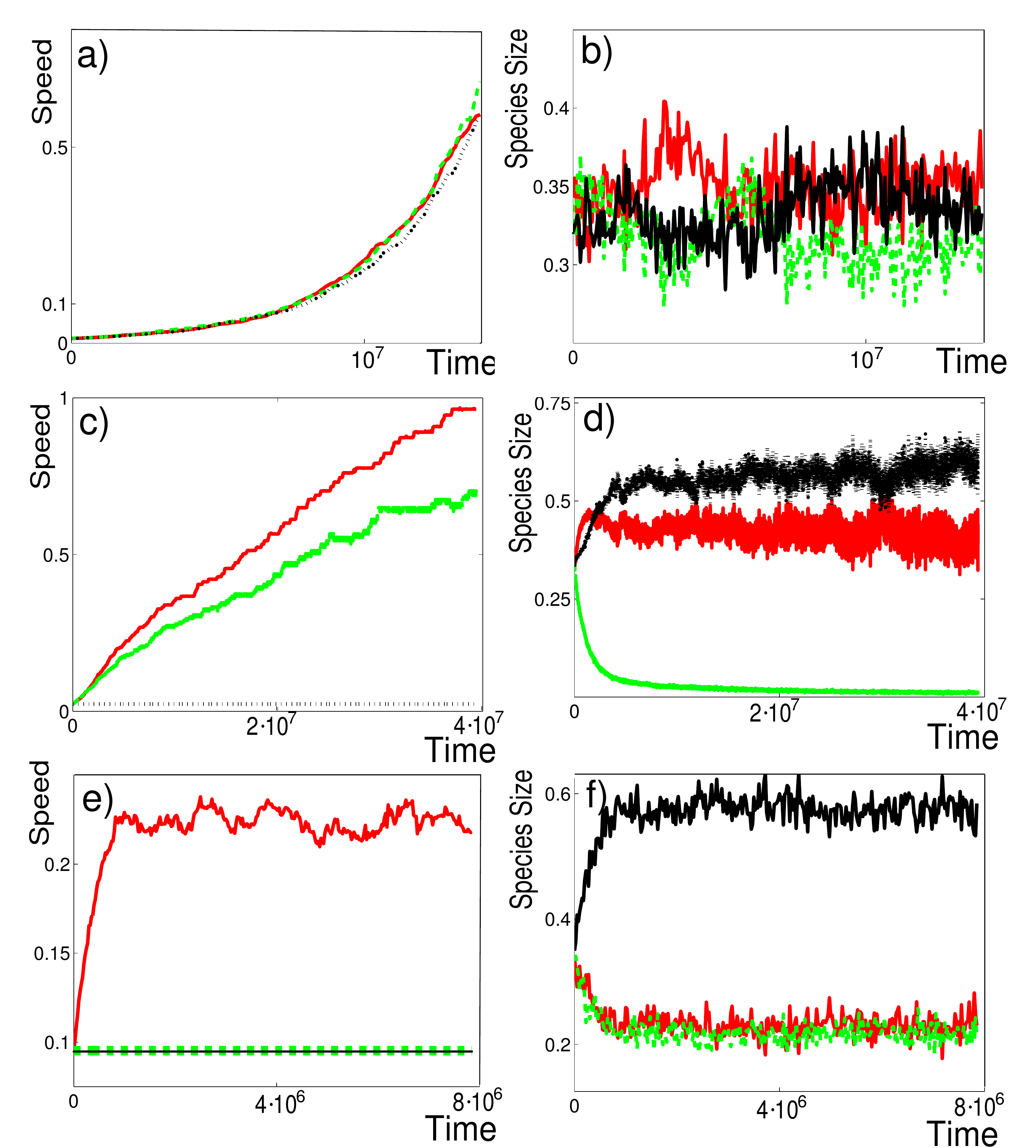}
            	\caption{(Color online)
                    	\textbf{a-b)} When all species mutate at the same rate, their growth rates will increase exponentially. They will, therefore, stay equally abundant.
                    	\textbf{c-d)} When only species 1 and 2 are allowed to mutate, their growth rates will steadily accelerate relative to species 3. Thus, species 2 will become scarcer on the lattice, until it dies out quickly followed by species 1. Simulations are carried out on a 1000x1000 lattice to better show the hyperbolic decline of species 2.
                    	\textbf{e-f)} When only species 1 is allowed to mutate, it accelerates to a growth rate fluctuating around 2.4 times faster than species 2 and 3. Consequently, species 3 grows to become 2.6 times more abundant than species 1 and 2.
            	}
            	\label{fig:speedCurves}
    	\end{center}
\end{figure}

If all three species are allowed to mutate, their growth rates will accelerate exponentially (see fig. \ref{fig:speedCurves}a). The growth rates for the three species stay approximately equal, so from \eqref{eq:speciesCounts} the species will remain equally abundant, as observed in fig. \ref{fig:speedCurves}b.

When both species 1 and 2 are allowed to mutate, both their growth rates appear to accelerate linearly. Thus, from \eqref{eq:speciesCounts}, species 1 and 3 will continue to grow while species 2 decreases hyperbolically in size until it goes exctint (see fig. \ref{fig:speedCurves}c-\ref{fig:speedCurves}d). After this, species 1 will quickly be overgrown by species 3, so only the non-mutating species is left. Due to the hyperbolic decrease in the abundance of species 2, the time before biodiversity is lost grows drastically with the size of the system.

When only species 1 is allowed to mutate, it will evolve to a certain mean growth rate $v_1 = (2.4 \pm 0.1) v_2$. At the same time, species 3 will grow in size to $p_3 = (2.6 \pm 0.1)p_1$, while the relative sizes of species 1 and 2 will remain about equal (see fig. \ref{fig:speedCurves}e-\ref{fig:speedCurves}f). The reason why $v_1 / v_2$ differs from $p_3 / p_1$ is that the mean field approximation \eqref{eq:speciesCounts} is not perfect. For larger system sizes the same mean growth rates are observed, but with a smaller variance.

Interestingly, biodiversity is maintained in our system when either one or three species are allowed to mutate, but not if two species are. 

\begin{figure}[tbp]
    	\begin{center}
                    	\includegraphics[width=\columnwidth]{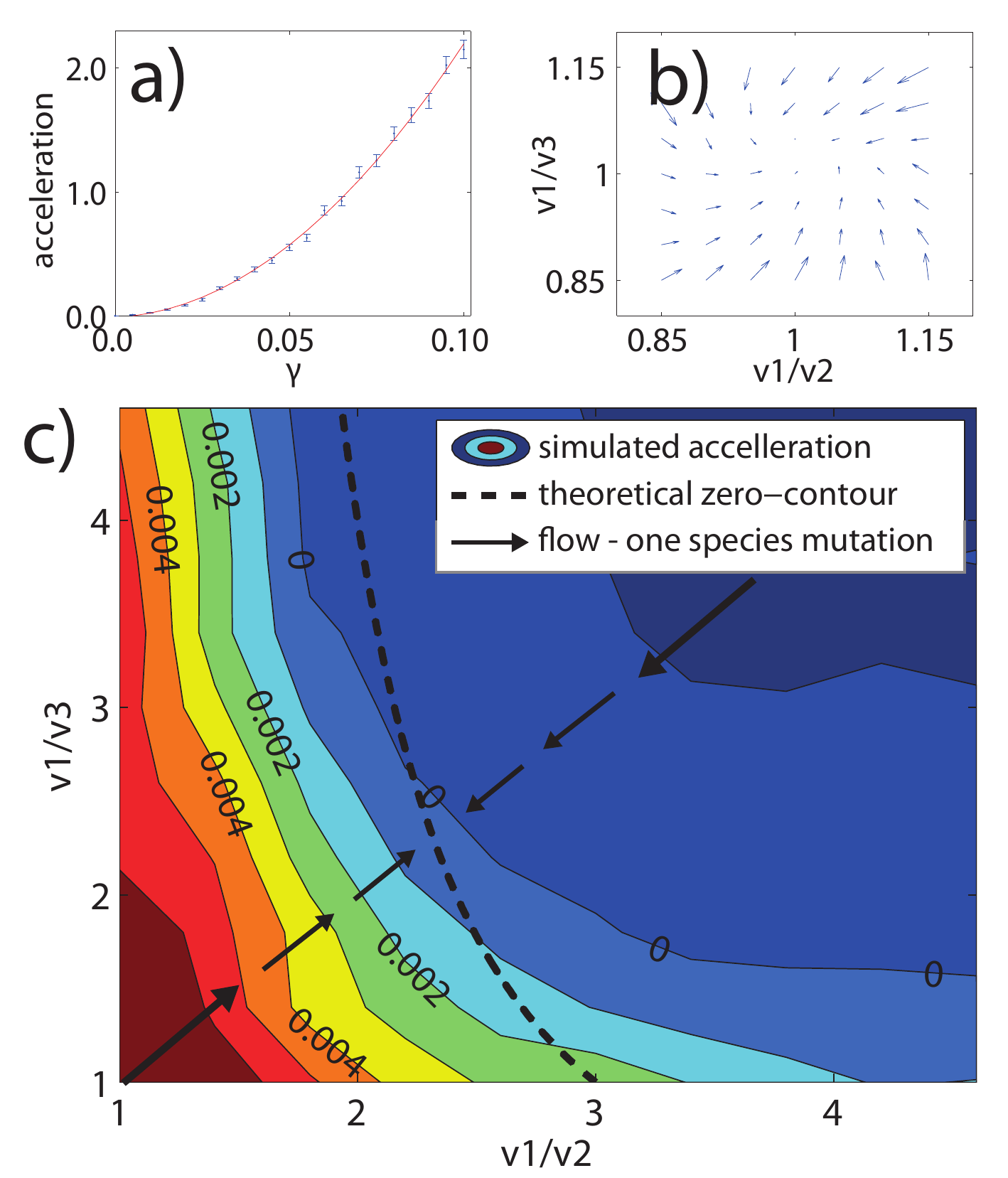}
            	\caption{(Color online)
                    	\textbf{a)} The acceleration in growth rate is proportional to the $\gamma$ squared. Thus, the evolution of species in cyclic competition communities will be dominated by big leaps in growth rate.
                    	\textbf{b)} When all species are allowed to mutate, all species having equal growth rate is a stable fixed point of the dynamics. If species 1 is growing faster than 2 and 3, faster mutants of species 1 will die out more frequently, thereby decreasing the growth rate.
                    	\textbf{c)} The relative acceleration in growth rate  $\frac{1}{v_1} \frac{d v_1}{d t}$ as a function of the relative velocities $v_1 / v_2$ and $v_1 / v_3$. When all species grow at the same rate, their growth rates will all accelerate. If a species is growing at a much higher rate than the two others, slower mutants will have better survival chances than faster mutants, thereby decreasing the growth rate. Arrows indicate the flow of the system, when only one species is allowed to mutate. The thick line marks the theoretically optimal speed of species 1, given by equation \eqref{eq:steadyState}. It is seen to agree well with observations for $v_2\approx v_3$.
            	}
            	\label{fig:phasespace}
    	\end{center}
\end{figure}

From \eqref{eq:speciesCounts} we get that the difference in growth rate between the fastest and the slowest species must stay bounded in order for the system to be stable. If not, the predator of the slowest species will decrease in size until it goes extinct. We therefore investigate the relative  acceleration $\frac{1}{v_1} \frac{d v_1}{d t}$ of the fastest species in the system. This is done by measuring how often a faster-growing mutant survives and, therefore, contributes to the total growth rate of the species. Naturally, this acceleration is proportional to the probability of mutation $p_{mutate}$, when this is small. More surprisingly, the acceleration is proportional to the square of $\gamma$ (see fig. \ref{fig:phasespace}a). This is due to the fact that both the survival chances of a mutant and the corresponding increase in growth rate are proportional to $\gamma$. The implications are that, for an ecological system where mutations of a broad range of magnitudes are expected to occur, the evolution of species will be dominated by big leaps. Small evolutionary improvements in fitness will most likely come to nothing.

In fig. \ref{fig:phasespace}c, the relative acceleration of the fastest species is plotted as a function of the relative growth rates $v_1/v_2$ and $v_1/v_3$. These have been found by monitoring how often fast mutants will succeed in outcompeting the species of growth rate $v_1$, compared to slow mutants. Each point is an average over the introduction of a minimum of 80000 mutants. It is seen that when the relative growth rates are large, the fastest species will decelerate. This is due to the better survival chances of the slow mutants decreasing the growth rate of the species and promoting biodiversity. Multiplying all growth rates by a constant factor corresponds to changing the time scales, so the functional form of the figure is independent of the absolute magnitudes of the growth rates.

When all three species are allowed to mutate, their growth rates will remain equal to each other. If a species becomes faster than the others, its acceleration will decrease, allowing the others to catch up (see fig. \ref{fig:phasespace}b and \ref{fig:phasespace}c). This constant increase in relative growth rate explains the exponential acceleration in fig. \ref{fig:speedCurves}a.

The case of two species mutating at the same rate corresponds in fig. \ref{fig:phasespace}c to one species moving along the horizontal line $v_1 / v_2=1$ and the other moving along the vertical line $v_1 / v_3 =1$. Along both lines, the acceleration is positive, so the system will mutate to extinction as seen in fig. \ref{fig:speedCurves}c. If one of the mutating species becomes faster than the other, it will decrease its acceleration, thus allowing the other to catch up.

Letting only one species mutate corresponds to moving along the line $v_1 / v_2 = v_1 / v_3$ in fig.  \ref{fig:phasespace}c until the acceleration becomes zero at a value just below 2.5, as indicated by the arrows. This explains fig. \ref{fig:speedCurves}e.

\textbf{Mechanism for deceleration}
---As demonstrated, the deceleration resulting from large relative growth rates is crucial for maintaining biodiversity. Using a simple, one-dimensional argument, we now derive a relation for the relative growth rates at steady state.

\begin{figure}[tbp]
    	\centering
            	\includegraphics[width=\columnwidth]{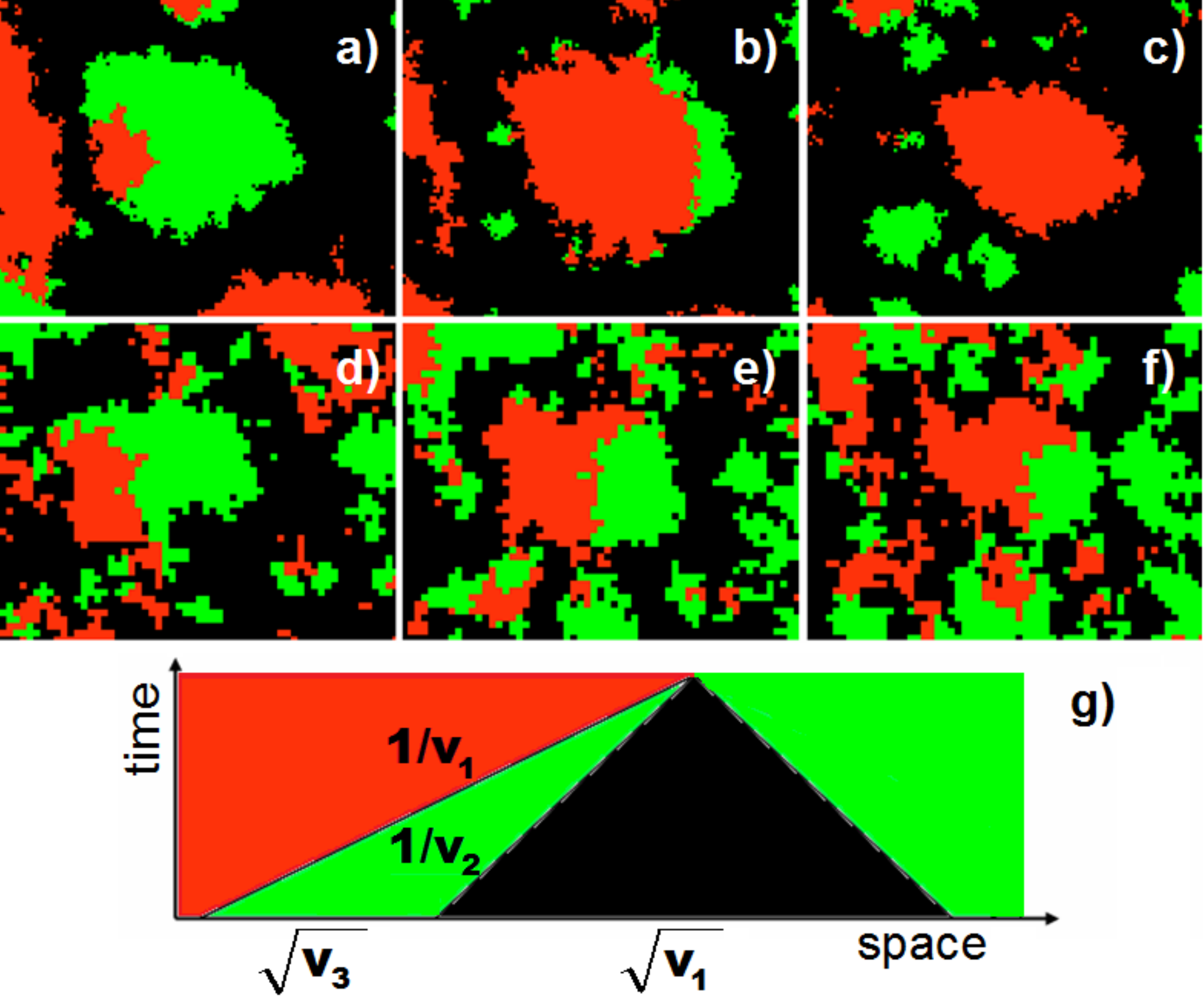}
    	\caption{(Color online)
            	\textbf{a-c)} Typical situation when species 1 is five times faster than species 2 and 3. Species 1 invades a cluster of species 2 and quickly overgrows it all, leaving it surrounded by its predator, species 3.
            	\textbf{d-f)} Typical situation when species 1 is 2.4 times faster than species 2 and 3. Before species 1 has overgrown all of species 2, this has grown through species 3 to connect to a new cluster of 2.
            	\textbf{g)} Clusters of species 1, 2, 3, and 2 arranged on a line. In the space-time diagram species 1 grows with slope $1/v_1$ and species 2 grows with slope $1/v_2$. A typical cluster length of species 2 is $\lambda_2 \propto \sqrt{v_3}$ and for species 3 it is $\lambda_3 \propto \sqrt{v_1}$. The optimal speed for species 1 follows by demanding that a typical cluster of species 2 is overgrown just when it connects to a new cluster of species 2.
    	}
    	\label{fig:optimalAndTooFastFull}
\end{figure}

Locally, a faster mutant will always have a competitive advantage over a slower mutant. However, mutants of different growth rates will have a tendency to separate spatially on the lattice \cite{Nahum}. The faster mutants are then at risk of exhausting their neighborhood of prey, leaving it in an isolated cluster surrounded by predators (see fig. \ref{fig:optimalAndTooFastFull}a-\ref{fig:optimalAndTooFastFull}c). To avoid this, the fastest species should allow time for its prey to grow through its predator, connecting it to a new cluster of prey (see fig.  \ref{fig:optimalAndTooFastFull}d-\ref{fig:optimalAndTooFastFull}f). At the optimal growth rate, a species will grow through a typical cluster of prey just when this becomes connected to a new cluster of prey.

Since the abundances of the three species are given by \eqref{eq:speciesCounts1}, one can expect the typical cluster sizes to scale like
\begin{eqnarray}
(\lambda_1,\lambda_2,\lambda_3) \propto (\sqrt{v_2},\sqrt{v_3},\sqrt{v_1}).
\label{eq:clusterSizes}
\end{eqnarray}

Imagine typical clusters of species 1, 2, 3, and 2 arranged on a one dimensional line (see fig. \ref{fig:optimalAndTooFastFull}g). As time passes, the two clusters of species 2 will overgrow species 3 at rate $v_2$. If species 1 is to overgrow the neighboring cluster of species 2 at the exact time when this becomes connected to the second cluster, we must have
\begin{eqnarray}
\frac{1}{v_1} \left( \sqrt{v_3} + \frac{\sqrt{v_1}}{2} \right) = \frac{1}{v_2} \frac{\sqrt{v_1}}{2} \Leftrightarrow \\
\frac{v_1}{v_2} = 1+ 2\left( \frac{v_1}{v_3} \right)^{-1/2}
\label{eq:steadyState}
\end{eqnarray}

When $v_2=v_3$, this has the solution $v_1 / v_2 = v_1 / v_3 = 2.3$, which is in excellent agreement with the value of $2.4 \pm 0.1$ found in fig. \ref{fig:speedCurves}e. In fig. \ref{fig:phasespace}c the contour line of zero acceleration is seen to agree well with \eqref{eq:steadyState} when $v_2 \approx v_3$. The discrepancies for $ v_2 \gg v_3$ and $ v_2 \ll v_3$ arise because the cluster structures of species 2 and 3 disappear in these limits, which negates \eqref{eq:clusterSizes} allowing species 1 to survive.

\textbf{Discussion}
---Our results explain quantitatively how communities of primitive organisms, such as bacteria and plants, in cyclic competition can evolve to a state with moderate consumption of a limited resource. Even though individuals would get a competitive advantage by growing fast, groups of fast growing individuals locally deplete their prey. Since individuals growing at different rates have a tendency to separate spatially, this group selection will favour moderate growth rates. Thus, the growth rate will be limited by the condition that clusters of prey should connect to other clusters of prey before being completely overgrown. The equation \eqref{eq:steadyState} describing this optimal growth rate is, to our knowledge, the first quantitative result in the field.

The spatial structure of the system is crucial for the group selection. In a well-mixed system the species will start a race to extinction. In a locally structured system, such as a Petri dish or the ocean bed, biodiversity can be maintained if one or all of the species are allowed to mutate. If two species are allowed to mutate, they will increase their growth rate until the system becomes unstable. This interesting result has not previously been reported.

If all species are allowed to mutate, but at different rates, the species with highest mutation rate will typically become eliminated after a long transient dynamics where all species increase their growth rates enormously. In practice this drastic increase in growth rates will be limited by metabolic constraints. Therefore, also systems with multiple evolving species should be stabilized before collapse.
In conclusion, our results emphasize the importance of modesty in growth, as well as modesty in the ability to evolve towards larger growth rates. In particular, in the case of cyclic competition our results quantitatively explain 
how groups of primitive organisms may self-organize to a state of sustainable development, preventing the tragedy of the commons.



\end{document}